\documentstyle[a4p,12pt,epsfig]{article}
\newcommand{\lumi}{55~{\rm pb}^{-1}}

\newcommand{\csrhi}{(0.9$\pm$0.3$\pm$0.1)~pb}
\newcommand{\csrlo}{(4.1$\pm$1.6$\pm$0.6)~pb}

\newcommand{\qqbar}{{\mathrm q}\bar{\mathrm q}}
\newcommand{\ffbar}{{\mathrm f}{\bar{\mathrm f}}}
\newcommand{\ee}{{\mathrm e}^+{\mathrm e}^-}
\newcommand{\ww}{{\rm W}^+{\rm W}^-}

\newcommand{\tautau}{\tau^+\tau^-}

\newcommand{\eeqq}{eeq$\bar{\mathrm q}$}

\newcommand{\Zg}{{\mathrm Z}/\gamma^{*}}
\newcommand{\Zee}{\mbox{$\ee \to$ (e)eZ}}
\newcommand{\Zgee}{\mbox{$\ee \to {\mathrm (e)e}\Zg$}}
\newcommand{\eeZg}{{\mathrm (e)e}\Zg}

\newcommand{\eetoww}{\ee\to\ww}

\newcommand{\eetogg}{\ee\to\ee+{\mathrm hadrons}}

\newcommand{\eetoqqbar}{\ee\to\qqbar}

\newcommand{\gev}{\mathrm GeV}

\newcommand{\ipb}{{\mathrm pb}^{-1}}

\newcommand{\dedx}{{\mathrm d}E/{\mathrm d}x}

\newcommand{\qe}{{\mathrm Q}_{\mathrm e}}
\newcommand{\qecosqq}{-\qe\cdot\cos\theta_{\qqbar}}
\newcommand{\qecose}{-\qe\cdot\cos\theta_{\mathrm e}}

\newcommand{\eelec}{E_{\rm e}}
\newcommand{\pelec}{p_{\rm e}}
\newcommand{\eop}{\eelec/\pelec}

\newcommand{\shat}{\hat{s}}
\newcommand{\that}{\hat{t}}
\newcommand{\uhat}{\hat{u}}

\newcommand{\mqq}{m_{\qqbar}}

\newcommand{\raiseb}{\raisebox{1.5ex}[-1.5ex]}

\newcommand{\Efwd}      {E_{\mathrm{fwd}}}
\newcommand{\pmiss}     {p_{\mathrm{miss}}}
\newcommand{\costmiss}  {\cos\theta_{\mathrm miss}}

\newcommand{\pet}       {p_{\mathrm{t}}}

\newcommand{\EPJ} {Eur.~Phys.~J.}
\newcommand{\PhysLett}  {Phys.~Lett.}

\newcommand{\PhysRev}   {Phys.~Rev.}
\newcommand{\ModPhys}  {Mod.~Phys.~Lett.}
\newcommand{\NPhys}  {Nucl.~Phys.}
\newcommand{\NIM} {Nucl.~Instr.~Meth.}
\newcommand{\CPC} {Comp.~Phys.~Comm.}
\newcommand{\ZPhys}  {Z.~Phys.}
\newcommand{\etal}     {et al.}
\newcommand{\OPALColl}  {OPAL Collaboration}

\begin{document}
\begin{titlepage}
\begin{center}{\large   EUROPEAN LABORATORY FOR PARTICLE PHYSICS
}\end{center}\bigskip
\begin{flushright}
       CERN-EP/98-120 \\ July 20, 1998
\end{flushright}
\bigskip\bigskip\bigskip\bigskip\bigskip
\begin{center}
{\huge\bf\boldmath
        First Measurement of $\Zg$\ Production 
        in Compton Scattering of Quasi-real
        Photons
}
\end{center}\bigskip\bigskip
\begin{center}{\LARGE The OPAL Collaboration
}\end{center}\bigskip\bigskip
\bigskip\begin{center}{\large  Abstract}\end{center}

We report the first observation of $\Zg$\ production in Compton
scattering of quasi-real photons.  This is a subprocess of the
reaction $\ee\to\ee\Zg$, where one of the final state electrons is
undetected. Approximately $\lumi$\ of data collected in the year 1997
at an $\ee$\ centre-of-mass energy of 183~GeV with the OPAL detector
at LEP have been analysed. The $\Zg$\ from Compton scattering has been
detected in the hadronic decay channel. Within well defined kinematic
bounds, we measure the product of cross-section and $\Zg$\ branching
ratio to hadrons to be \csrhi\ for events with a hadronic mass larger
than 60~GeV, dominated by (e)eZ production. In the hadronic mass
region between 5~GeV and 60~GeV, dominated by (e)e$\gamma^*$
production, this product is found to be \csrlo. Our results agree
with the predictions of two Monte Carlo event generators, grc4f and
PYTHIA.

\bigskip\bigskip\bigskip\bigskip
\bigskip\bigskip
\begin{center}{\large
(Submitted to Physics Letters B)
}\end{center}
\end{titlepage}
\begin{center}{\Large        The OPAL Collaboration
}\end{center}\bigskip

\begin{center}{
G.\thinspace Abbiendi$^{  2}$,
K.\thinspace Ackerstaff$^{  8}$,
G.\thinspace Alexander$^{ 23}$,
J.\thinspace Allison$^{ 16}$,
N.\thinspace Altekamp$^{  5}$,
K.J.\thinspace Anderson$^{  9}$,
S.\thinspace Anderson$^{ 12}$,
S.\thinspace Arcelli$^{ 17}$,
S.\thinspace Asai$^{ 24}$,
S.F.\thinspace Ashby$^{  1}$
% ...
}
\end{center}\bigskip
\bigskip
$^{  1}$School of Physics and Astronomy, University of Birmingham,
Birmingham B15 2TT, UK
\newline
$^{  2}$Dipartimento di Fisica dell' Universit\`a di Bologna and INFN,
I-40126 Bologna, Italy
\newline
$^{  3}$Physikalisches Institut, Universit\"at Bonn,
D-53115 Bonn, Germany
\newline
$^{  4}$Department of Physics, University of California,
Riverside CA 92521, USA
\newline
$^{  5}$Cavendish Laboratory, Cambridge CB3 0HE, UK
\newline
$^{  6}$Ottawa-Carleton Institute for Physics,
Department of Physics, Carleton University,
Ottawa, Ontario K1S 5B6, Canada
\newline
$^{  7}$Centre for Research in Particle Physics,
Carleton University, Ottawa, Ontario K1S 5B6, Canada
\newline
$^{  8}$CERN, European Organisation for Particle Physics,
CH-1211 Geneva 23, Switzerland
\newline
$^{  9}$Enrico Fermi Institute and Department of Physics,
University of Chicago, Chicago IL 60637, USA
\newline
$^{ 10}$Fakult\"at f\"ur Physik, Albert Ludwigs Universit\"at,
D-79104 Freiburg, Germany
\newline
$^{ 11}$Physikalisches Institut, Universit\"at
Heidelberg, D-69120 Heidelberg, Germany
\newline
$^{ 12}$Indiana University, Department of Physics,
Swain Hall West 117, Bloomington IN 47405, USA
\newline
$^{ 13}$Queen Mary and Westfield College, University of London,
London E1 4NS, UK
\newline
$^{ 14}$Technische Hochschule Aachen, III Physikalisches Institut,
Sommerfeldstrasse 26-28, D-52056 Aachen, Germany
\newline
$^{ 15}$University College London, London WC1E 6BT, UK
\newline
$^{ 16}$Department of Physics, Schuster Laboratory, The University,
Manchester M13 9PL, UK
\newline
$^{ 17}$Department of Physics, University of Maryland,
College Park, MD 20742, USA
\newline
$^{ 18}$Laboratoire de Physique Nucl\'eaire, Universit\'e de Montr\'eal,
Montr\'eal, Quebec H3C 3J7, Canada
\newline
$^{ 19}$University of Oregon, Department of Physics, Eugene
OR 97403, USA
\newline
$^{ 20}$CLRC Rutherford Appleton Laboratory, Chilton,
Didcot, Oxfordshire OX11 0QX, UK
\newline
$^{ 22}$Department of Physics, Technion-Israel Institute of
Technology, Haifa 32000, Israel
\newline
$^{ 23}$Department of Physics and Astronomy, Tel Aviv University,
Tel Aviv 69978, Israel
\newline
$^{ 24}$International Centre for Elementary Particle Physics and
Department of Physics, University of Tokyo, Tokyo 113, and
Kobe University, Kobe 657, Japan
\newline
$^{ 25}$Institute of Physical and Environmental Sciences,
Brunel University, Uxbridge, Middlesex UB8 3PH, UK
\newline
$^{ 26}$Particle Physics Department, Weizmann Institute of Science,
Rehovot 76100, Israel
\newline
$^{ 27}$Universit\"at Hamburg/DESY, II Institut f\"ur Experimental
Physik, Notkestrasse 85, D-22607 Hamburg, Germany
\newline
$^{ 28}$University of Victoria, Department of Physics, P O Box 3055,
Victoria BC V8W 3P6, Canada
\newline
$^{ 29}$University of British Columbia, Department of Physics,
Vancouver BC V6T 1Z1, Canada
\newline
$^{ 30}$University of Alberta,  Department of Physics,
Edmonton AB T6G 2J1, Canada
\newline
$^{ 31}$Duke University, Dept of Physics,
Durham, NC 27708-0305, USA
\newline
$^{ 32}$Research Institute for Particle and Nuclear Physics,
H-1525 Budapest, P O  Box 49, Hungary
\newline
$^{ 33}$Institute of Nuclear Research,
H-4001 Debrecen, P O  Box 51, Hungary
\newline
$^{ 34}$Ludwigs-Maximilians-Universit\"at M\"unchen,
Sektion Physik, Am Coulombwall 1, D-85748 Garching, Germany
\newline
\bigskip\newline
$^{  a}$ and at TRIUMF, Vancouver, Canada V6T 2A3
\newline
$^{  b}$ and Royal Society University Research Fellow
\newline
$^{  c}$ and Institute of Nuclear Research, Debrecen, Hungary
\newline
$^{  d}$ and Department of Experimental Physics, Lajos Kossuth
University, Debrecen, Hungary
\newline
$^{  e}$ on leave of absence from the University of Freiburg
\newline

\section{Introduction}
\label{introduction}

\begin{figure}[h] 
\center{
\epsfig{file=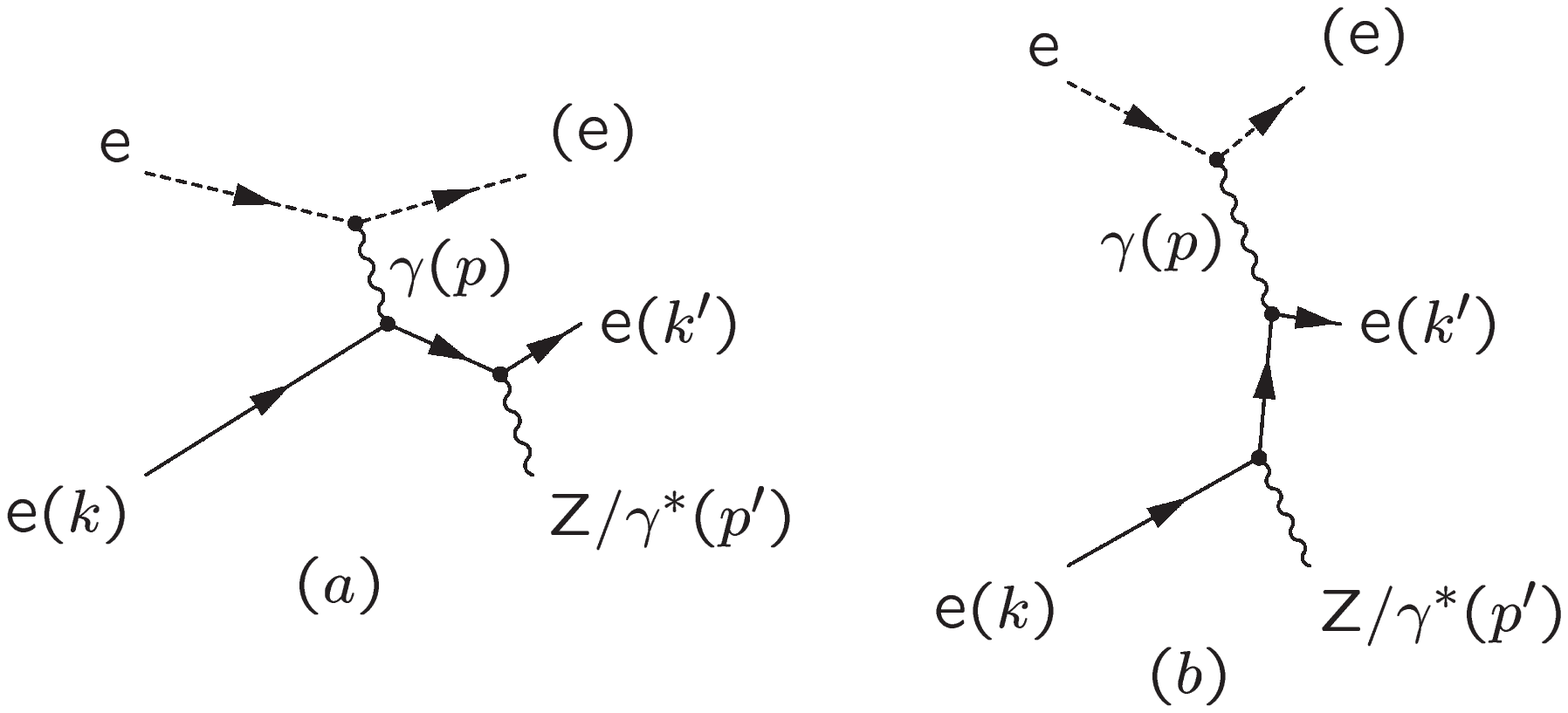,width=0.8\textwidth}
\caption{\it Diagrams for the process \Zgee.}
\label{fig:zee} 
}
\end{figure}

Single Z boson production in $\ee$ collisions~\cite{SINGLEZ},
\Zee, will be the dominant source of Z bosons at linear $\ee$ 
colliders with centre-of-mass energies above
500~GeV~\cite{HAGIWARA91}.  At LEP2, however, the cross-section is
nearly two orders of magnitude below the ``radiative return'' process
$\ee\to\gamma$Z.  The elementary subprocess of single Z boson
production is\footnote{Charge conjugation is implied throughout the
paper except when otherwise stated.} \mbox{e$^-\gamma \to$ e$^-$Z}
where a quasi-real photon radiated from one of the beam electrons
scatters off the other electron producing a Z as shown in
Fig.~\ref{fig:zee}.  This process is ordinary Compton scattering with
the outgoing real photon replaced by a Z. Its cross-section has first
been calculated for ep collisions~\cite{EPSINGLEZ}, where the incoming
photon is radiated from the proton. Another variant of Compton
scattering is \mbox{e$^-\gamma \to$ e$^-\gamma^*$,} where this time
the outgoing real photon is replaced by a virtual one.  The observable
final state (e)e$\ffbar$\ will in both cases be the scattered electron
e and a fermion pair $\ffbar$\ from the $\Zg$\ decay, while the other
electron (e) usually remains unobserved in the beam pipe.

The final state (e)e$\ffbar$\ may contribute an important background
to many searches for new physics, for example final states involving
large missing masses ($\ffbar=\nu\bar{\nu}$), leptoquark production in
deep inelastic electron-photon scattering ($\ffbar=\qqbar$), new
particles decaying to Z bosons, and excited electrons in the decay
channel e$\ffbar$.  Excited electrons can for example directly
contribute in the $\hat{s}$ and $\hat{u}$ channel diagrams of the
signal in Fig.~\ref{fig:zee}.

For Z boson production in Compton scattering e$(k)\gamma(p) \to$
e$(k^\prime)$Z$(p^\prime)$ of real photons ($p^2~=~0$), the cross-section
dependence on the Mandelstam variables
$\hat{s}=(k^\prime+p^\prime)^2$, $\hat{t}=(k^\prime-k)^2$,
$\hat{u}=(p^\prime-k)^2$ is~\cite{EPSINGLEZ}
\begin{equation}
\frac{{\rm d}\sigma}{{\rm d}\hat{t}} \propto
\frac{1}{\hat{s}^2}\left(
        \frac{\hat{u}}{\hat{s}} +
        \frac{2M_{\rm Z}^2\hat{t}}{\hat{u}\hat{s}} +
        \frac{\hat{s}}{\hat{u}} \right).
\label{eq:compton}
\end{equation} 
For $M_{\rm Z}= 0$ the well-known terms for ordinary Compton
scattering remain. The typical transverse momentum scale of the
scattered $\Zg$\ bosons is small, leading to a singularity at
$\hat{u}=0$ of the virtual electron propagator in
Fig.~\ref{fig:zee}b. For incoming quasi-real photons ($p^2\approx~0$)
in ep or $\ee$ collisions, the dominant regulating effect for this
divergence is not the electron mass, but small, non-zero, incoming
photon masses squared $p^2$, which introduce via the
replacement~\cite{EPSINGLEZ}
\begin{equation}
 \hat{u} \to \hat{u} + p^2\frac{M_{\rm Z}^2}{\hat{s}}
\label{eq:regulate}
\end{equation}
a lower cutoff in the denominator of Eq.~(\ref{eq:compton}). A simple
equivalent photon approximation (EPA), as used in
Ref.~\cite{FORMEREP}, where the integration over the range of small
photon virtualities leads to an effective on-shell incoming photon
flux, overestimates the cross-section by about a factor of
two~\cite{HAGIWARA91}. In order to properly describe the process, the
$p^2$ spectrum of the incoming photons has to be retained fully, or a
modified EPA~\cite{HAGIWARA91} has to be introduced. In any case the
results will be sensitive to the modelling of the $p^2$ spectrum.  In
this paper the theoretical expectations are represented by Monte Carlo
event generators which use different approaches for obtaining the
$p^2$ spectrum of the incoming photons. These are compared with
experimental data for the first time.

We have analysed the process \Zgee\ using data collected with the OPAL
detector at an $\ee$\ centre-of-mass energy of about 183~GeV,
corresponding to a total integrated luminosity of approximately
$\lumi$. The Z or the $\gamma^*$\ is observed via its decay to
hadrons.

\section{The OPAL detector}
\label{opaldet}

A detailed description of the OPAL detector can be found
elsewhere~\cite{opaldetector}. Subdetectors which are particularly
relevant for the present analysis are described here briefly.  The
central detector consists of a system of tracking chambers providing
charged particle tracking over 96\% of the full solid angle inside a
0.435~T uniform magnetic field parallel to the beam axis\footnote{The
OPAL coordinate system is a right-handed system which is defined such
that the $z$-axis is in the direction of the electron beam and the
$x$-axis is horizontal and points towards the centre of the LEP ring;
$\theta$ is the polar angle with respect to $z$ and $\phi$ is the
azimuthal angle about the $z$ axis.}. Starting with the innermost
components, it consists of a high precision silicon microvertex
detector, a precision vertex drift chamber, a large volume jet chamber
with 159 layers of axial anode wires and a set of $z$ chambers
measuring the track coordinates along the beam direction. With the jet
chamber, the momenta of tracks can be measured with an accuracy of
$\sigma_p/p^2 = 2 \times 10^{-3}~\gev^{-1}$.  The jet chamber also
provides energy loss measurements which are used for particle
identification.

A lead-glass electromagnetic calorimeter (ECAL) located outside the
magnet coil covers the full azimuthal range with excellent hermeticity
in the polar angle range of $|\cos \theta |<0.82$ for the barrel
region and $0.81<|\cos \theta |<0.984$ for the endcap region.  The
forward detectors and silicon tungsten calorimeters~\cite{opalsw}
located at both sides of the interaction point measure the integrated
luminosity and complete the geometrical acceptance down to 33~mrad in
polar angle.  The magnet return yoke is instrumented with streamer
tubes for hadron calorimetry and is surrounded by several layers of
muon chambers.

\section{Event simulation}
\label{evsimulation}

The \Zgee\ signal events are generated using the grc4f Monte Carlo
program~\cite{grc4f}.  Hadronization is performed by
JETSET~\cite{jetset}. A sample of approximately 170 times the
integrated luminosity of the data was used. In grc4f, an automatic
computation system for Feynman diagrams, GRACE~\cite{GRACE} is
invoked, yielding the differential cross-sections of tree-level
diagrams by performing a calculation of the matrix elements and a
Monte Carlo integration over the phase space. Initial state radiation
corrections are also implemented.  The delicate $p^2$ distribution of
the incoming photons for the e$\gamma$ Compton scattering is thus
obtained by a numerical integration of the matrix element.

As a second Monte Carlo event generator for the process \Zgee\ we used
the PYTHIA~\cite{pythia} program. To include the full phase space for
this low-$\pet$ process, we lowered the $\widehat{\pet}$ cutoff in the
e$\gamma$ rest frame from the default of 1~GeV to 10~MeV in the
generation of the events, as suggested in the PYTHIA manual. Without
this change PYTHIA would underestimate largely the cross-section. In
PYTHIA, the signal process is simulated according to
Eqs.~(\ref{eq:compton}) and (\ref{eq:regulate}). An equivalent flow of
incoming photons is modelled with a $p^2$ distribution obtained from a
first order QED expression~\cite{pythia}.  This approximation
implemented in PYTHIA is used as a cross-check, with a sample size of
about 1/10 of that generated with grc4f.

Backgrounds from other four-fermion final states have been studied
using several samples also generated with grc4f and JETSET. Background
contributions from two-photon processes, $\eetogg$, were evaluated
using PHOJET~\cite{phojet} for events with low $Q^2$ and
HERWIG~\cite{herwig} for high-$Q^2$ processes. Double counting of the
\eeqq\  final state between signal, two-photon and other four-fermion
samples is avoided by separating the contributions from different
classes of diagrams, or by performing kinematic cuts. Two-photon
production of $\tautau$\ was modelled with the
VERMASEREN~\cite{vermaseren} generator. Other background processes
involving two fermions in the final state were simulated using PYTHIA
for the channel $\ee\to\qqbar$ and KORALZ~\cite{koralz} for the
channel $\ee\to\tautau$. Background contributions from other processes
have been checked and found to be negligible.

All Monte Carlo events were passed through the full simulation of the
OPAL detector~\cite{gopal} and then subjected to the same
reconstruction and analysis procedures as the real data.

\section{Signal definition}
\label{sigdef}

In most of the \Zgee\ events the first incoming electron, which
radiates the photon taking part in the Compton process, remains
undetected in the beam pipe due to the low momentum transfer $p^2$.
In contrast, the second electron is often emitted at large angles.
Therefore the predominant signature of the process is a single
scattered electron isolated from a hadronic system containing one or
two hadron jets from the $\Zg$\ decay.  The differential distribution
is peaked at low $p^{\prime 2}$ (low mass $\qqbar$\ systems) and at
low $|\uhat|$ where the outgoing $\Zg$ is emitted almost in the
direction of the second incoming electron. Since for low $p^{\prime
2}$ and low $|\hat{u}|$ a huge part of the cross-section remains
unobserved in the beam pipe, it is not practical to measure the signal
process \Zgee\ in the whole phase space corresponding to the diagrams
in Fig.~\ref{fig:zee}.  Furthermore, considerable uncertainties due to
hadronic resonances for low mass $\gamma^*$ provide a supplementary
reason for limiting the phase space. The $\Zgee$\ signal process is
therefore defined by the diagrams in Fig.~\ref{fig:zee} and by
applying the following additional kinematic cuts on the \eeqq\ final
state.

\begin{itemize}
\item
One of the primary electrons is required to remain in the beam pipe:
$|\cos\theta_{\rm e_1}|>0.9995$. This angular cut is chosen close
to the edge of the 
silicon tungsten calorimeter.
\item
The second beam electron has to form a large angle with respect to the
beam: $|\cos\theta_{\rm e_2}|<0.985$, corresponding to the edge of
the electromagnetic endcap calorimeter.
\item
In order to detect the hadronic final state from the $\Zg$ decay, we
demand at least one primary quark with $|\cos\theta_{\rm q}|<0.985$.
\item
We require the invariant mass of the primary quark pair to be greater
than 5~GeV.
\end{itemize}

The cross-section times branching ratio predicted for $\Zgee$\ events
with an \eeqq\ final state satisfying the signal definition has been
evaluated using the grc4f and PYTHIA programs at an $\ee$\
centre-of-mass energy of 183~GeV. We obtain values of
(4.80$\pm$0.02)~pb for grc4f and (4.76$\pm$0.04)~pb for PYTHIA.

In addition to the e$\gamma$ Compton scattering in the process
$\Zgee$, there are two other classes of processes which lead to an
\eeqq\  final state, namely $\ee$\ annihilation, dominated by $\Zg -
\Zg$\ intermediate states, and two-photon interactions.  Of the former
process, only a negligible cross-section of 0.002~pb fulfils the
kinematic cuts of our signal definition, since the topology with
exactly one electron very close to the beam direction and the other
electron visible at large angles is not preferred.  Among the
two-photon interactions, on the other hand, exactly this
``single-tagged'' topology is typical for deep inelastic e$\gamma$
scattering with large momentum transfer squared\footnote{Note, that
$q^2$ in two-photon processes is by definition identical to $\that$\
in our signal process.},  $q^2 \equiv -Q^2$. According to PHOJET and
HERWIG, two-photon events which satisfy the above kinematic cuts are
produced with a cross-section of approximately 5.6~pb.

Using grc4f we checked that possible interferences between the
different classes of Feynman diagrams contributing to the \eeqq\ final
state, e$\gamma$ Compton scattering, $\ee$\ annihilation, and
two-photon interaction, are negligible. It is therefore justified to
treat each process separately and to add the predictions of the
respective Monte Carlo generators. Only $\Zgee$\ events which fulfil
the signal definition are treated as ``signal''.  All other processes
leading to an \eeqq\ final state are treated as ``background'' even if
the kinematic signal definition is fulfilled.  Likewise, $\Zgee$\
events that do not fulfil the signal definition are treated as
``background''.

\section{Event preselection}
\label{preselection}

We perform a preselection of events with a hadronic final state and
an isolated electron candidate.  Only tracks and clusters which
satisfy standard quality criteria are considered in the
analysis. The following preselection cuts are applied.
\begin{itemize}
\item 
We reject low multiplicity events by requiring the sum $N_{\rm mult}$
of the number of tracks and electromagnetic clusters not associated 
with any track to be greater than 8.
\item 
We search for electron candidates among all tracks using the following
criteria. We require the probability of the measured ionization energy
loss $\dedx$ of the track being consistent with an electron hypothesis
to be greater than 5\%.  Furthermore we require $\eop > 0.6$ where
$\eelec$ is the total energy in the electromagnetic calorimeter
associated with the track and $\pelec$ is the track momentum.  For all
electron candidate tracks we define a cone energy $E_{\rm cone}$ as
the sum of all track momenta and electromagnetic cluster energies in a
cone with a half angle of 0.2~rad around the track, excluding the
momentum and the cluster energy associated to the electron candidate
track. If more than one candidate is found in an event, the one which
has the smallest $E_{\rm cone}$ is chosen as the signal electron.  To
suppress events with fake electron candidates, we require the angle
between the signal electron and the closest track to be greater than
0.25~rad and $\eelec$ to be larger than $2~\gev$.
\item 
In order to be consistent with the signal definition, we require the
invariant mass of the $\qqbar$\ system $\mqq$\ to be larger than
5~GeV.  This mass is calculated from the momentum of all tracks and
the energy deposited in the electromagnetic and hadronic calorimeters,
the forward detectors and silicon tungsten calorimeters, excluding
those associated with the signal electron, using an
algorithm~\cite{gce} which corrects for double counting of energy
between tracks and calorimeter clusters.
\end{itemize}

In Fig.~\ref{fig:plotpres} we compare the measured distributions of
$\mqq$\ and $\eelec$ after preselection with the corresponding Monte
Carlo expectations. The gross features of the data are in satisfactory
agreement with the prediction at this early stage of the analysis.
Discrepancies, especially at low values of $\mqq$\ and $E_{\rm e}$,
can be attributed to systematic uncertainties in the two-photon
process $\eetogg$\ which is difficult to model in this kinematic
region. The apparent excess of Monte Carlo events in the range
105~GeV~$< \mqq < 160$~GeV has no influence on the analysis, since it
is populated nearly exclusively by background events, which will be
rejected later on.  A total of 1558$\pm$25 events is expected, where
the main contributions are from two-photon interactions, $\eetoww$\
and multihadronic events. In the data 1882 events are found after the
preselection (see Table~\ref{tab:evnum}).  As seen from
Fig.~\ref{fig:plotpres}, the data excess is located in kinematic
regions where the two-photon background dominates.
 
The expected signal contribution is 38.0$\pm$0.5 \Zgee\ events,
corresponding to an average signal efficiency of (14.5$\pm$0.2)\%
obtained with the grc4f Monte Carlo program. The corresponding value
obtained from PYTHIA is (16.7$\pm$0.6)\%.  A large part of the
preselection inefficiency for all $\Zg$\ masses is due to the typical
small transverse momenta of the electron and the $\Zg$\ in the final
state.  Signal electrons which form large angles with respect to the
beam therefore often have momenta below 2~GeV.  In addition, for low
$\gamma^*$\ masses, the $\gamma^*$\ decay products tend to be in a
rather narrow cone preferring forward directions, where particles may
escape undetected along the beam pipe. As a result many of these
events fail the multiplicity cut. This reflects in a significantly
lower preselection efficiency for smaller hadronic masses. Considering
only signal events with $ \mqq <$~60~GeV, grc4f predicts the
preselection efficiency to be (9.5$\pm$0.2)\%, while for $ \mqq
>$~60~GeV the efficiency is (50.3$\pm$0.7)\%.  The PYTHIA prediction
is consistent with these numbers.

\begin{table}[htb]
\begin{center} 
{\footnotesize
\begin{tabular}{|l||r|r|r|r|r|r|c|} \hline
&\multicolumn{6}{|c|}{Number of expected events from MC}& OPAL \\ 

\raiseb{Cut}
&\multicolumn{1}{|c}{$\eeZg$} 
&\multicolumn{1}{c}{$\gamma\gamma$} 
&\multicolumn{1}{c}{$\qqbar$}
&\multicolumn{1}{c}{4f} 
&\multicolumn{1}{c}{other}
&\multicolumn{1}{c|}{Sum}  
&\multicolumn{1}{|c|}{data} \\ \hline \hline

Preselection
& {38.0$\pm$0.5}
& {1251.9$\pm$23.9}    
& {113.8$\pm$3.6}    
& {143.1$\pm$0.9}    
& {11.6$\pm$0.6}    
& {1558.4$\pm$24.7}    
& {1882} \\
\hline

$\qecosqq > 0.75$
& {33.0$\pm$0.4}    
& {575.0$\pm$16.5}    
& {42.6$\pm$2.2}    
& {11.6$\pm$0.3}    
& {3.5$\pm$0.4}    
& {665.7$\pm$16.7}    
& {766} \\
\hline

$\qecose < 0.75$
& {26.5$\pm$0.4}    
& {209.5$\pm$10.1}    
& {30.8$\pm$1.9}    
& {9.3$\pm$0.2}    
& {2.0$\pm$0.3}    
& {278.1$\pm$10.3}    
& {294} \\
\hline

$|\costmiss| > 0.95$
& {24.3$\pm$0.4}    
& {128.9$\pm$7.8}    
& {20.2$\pm$1.5}    
& {2.9$\pm$0.1}    
& {1.1$\pm$0.2}    
& {177.5$\pm$7.9}    
& {193} \\
\hline

$\pmiss > 30$~GeV
& {21.6$\pm$0.3}    
& {21.0$\pm$2.0}    
& {15.9$\pm$1.4}    
& {2.0$\pm$0.1}    
& {0.7$\pm$0.1}    
& {61.2$\pm$2.5}    
& {74} \\
\hline

electron isolation
& {17.5$\pm$0.3}    
& {8.2$\pm$1.2}    
& {1.9$\pm$0.5}    
& {1.4$\pm$0.1}    
& {0.6$\pm$0.1}    
& {29.6$\pm$1.3}    
& {36} \\
\hline

$\Efwd < 35$~GeV
& {16.7$\pm$0.3}    
& {3.1$\pm$0.8}    
& {1.8$\pm$0.5}    
& {1.4$\pm$0.1}    
& {0.3$\pm$0.1}    
& {23.2$\pm$0.9}    
& {27} \\
\hline

\end{tabular} 
}
\caption{\it Numbers of expected and observed events for a luminosity
of $\lumi$\ after application of each cut. Only the most important
sources of background are shown separately. The number of expected
signal events is obtained using the grc4f generator. The numbers of
background events are evaluated using the Monte Carlo samples (MC)
described in the text. The errors are statistical only.}
\label{tab:evnum} 
\end{center} 
\end{table} 

\section{Selection of {\boldmath \Zgee} events}
\label{selection}

In order to reduce the remaining background six further cuts have been
applied which can be divided into three groups. The first group
consists the following angular cuts.
\begin{itemize}
\item 
In the $\Zgee$\ signal process, the $\Zg$\ is usually scattered very
close to the beam direction of the detected signal electron, given by
$-\qe \cdot \cos\theta=1$, where $\qe$\ is the charge (in units of
$e$) of the signal electron. We therefore require $\qecosqq >0.75$,
where $\cos\theta_{\qqbar}$ is the cosine of the polar angle of the
$\qqbar$\ system (see Fig.~\ref{fig:plotcuts}a).
\item 
To suppress the otherwise irreducible background of single-tagged
two-photon events, which fulfil the kinematic signal definition, we
take advantage of the fact that the distribution of the polar angle
$\cos\theta_e$ of the tagged electrons peaks around their beam
direction. We thus only retain events where $\qecose$\ is less than
0.75. This cut is shown in Fig.~\ref{fig:plotcuts}b. Note that
two-photon events with fake electrons, distributed symmetrically in
$\qecose$, largely dominate this spectrum. Correctly identified tagged
electrons of high $Q^2$ two-photon events are visible as an asymmetry
in the region near +1.
\end{itemize} 
Signal events are further characterised by an undetected electron
along the beam direction, so that cuts on the direction and amount of
the missing momentum in the event can be applied.
\begin{itemize}
\item 
We require the absolute value of the cosine of the polar angle of the
missing momentum $|\costmiss|$ to be larger than 0.95 (see
Fig.~\ref{fig:plotcuts}c). The largest relative suppression of this
cut occurs for $\eetoww$\ events, where one W boson decays to hadrons
and the other to an electron and a neutrino, since the neutrino is not
preferentially emitted close to the beam direction.
\item 
The distribution of the missing momentum after the cut on
$|\costmiss|$ is shown in Fig.~\ref{fig:plotcuts}d.
Requiring a missing momentum of at least $30~\gev$ largely suppresses
remaining background with fake or conversion electrons from untagged
two-photon events where both beam electrons stay under small angles
with respect to the beam axis and tend to balance the momentum.
\end{itemize}
The third group of cuts is applied to suppress events with fake
electron candidates.
\begin{itemize}
\item 
The minimum angle between the electron and the nearest charged track
was required to be 0.65~rad. This cut removes most of the remaining
multihadronic background as shown in Fig.~\ref{fig:plotcuts}e.
\item  
The sample still includes a small amount of two-photon background
tagged by an electron in the forward calorimeters where the signal
electron is fake or due to a double-tagged event topology.  It is
suppressed significantly by requiring the total energy deposit in the
forward detectors and silicon tungsten calorimeters, $\Efwd$, not to
exceed 35~GeV (see Fig.~\ref{fig:plotcuts}f).
\end{itemize}

The cuts, together with the expected and observed numbers of events, are
shown in Table~\ref{tab:evnum}.  Already after the second cut there
remains no significant excess in the data compared with the
expectation. The six selection cuts improve the signal to background
ratio by about a factor of 100 compared with the preselection, while
retaining about half of the signal events.

The efficiencies after all cuts are given in Table~\ref{tab:vglneu}.
Considering only signal events with $\mqq <$~60~GeV, dominated by
(e)e$\gamma^*$, grc4f predicts an overall selection efficiency of
(4.1$\pm$0.1)\%. For $\mqq>60$~GeV, dominated by (e)eZ, the overall
efficiency is (22.3$\pm$0.6)\%.  The corresponding numbers from
PYTHIA, (4.3$\pm$0.3)\% and (22.3$\pm$1.7)\%, are consistent with
grc4f within their statistical errors.  From the two lowest lines in
Table~\ref{tab:vglneu}, however, it is evident that PYTHIA predicts a
slightly different mix of high mass ``eeZ''-type and low mass
``ee$\gamma^*$''-type events than grc4f.  This causes about a 15\%
relative difference between the two Monte Carlo generators for
efficiencies averaged over the total $\mqq$ range and explains the
differences in the total preselection efficiencies given at the end of
Section~\ref{preselection}.  Since the predicted efficiency of each
single cut is consistent between the two simulations, this difference
remains essentially unchanged after the final selection cuts.

For calculating cross-sections we avoid assumptions on the mix of
``eeZ''-type, and ``ee$\gamma^*$''-type events. We instead calculate
the cross-sections separately, for masses above and below 60~GeV, the
point where the differential $\mqq$\ distribution has its minimum,
using the efficiencies obtained from grc4f.  After correcting for
feed-through from the low to the high mass region, and vice versa, we
obtain the results listed in Table~\ref{tab:vglneu}. The measured 
cross-sections agree with the predictions of the two Monte Carlo 
generators.

\begin{table}[htb]
\begin{center} 
\begin{tabular}{|l||c|c|} \hline
&$5~\gev < \mqq < 60~\gev$ & $\mqq \geq 60~\gev$              \\

                        &``ee$\gamma^*$''& ``eeZ''            \\ \hline \hline
efficiency (\%)         & 4.1$\pm$0.1 & 22.3$\pm$0.6 \\ \hline 

signal expected         & 9.5$\pm$0.2 & 7.2$\pm$0.2  \\
background              & 4.4$\pm$0.8 & 2.1$\pm$0.4  \\
feed-through from       &                &           \\ 
neighbouring mass region&\raiseb{0.2}    &\raiseb{0.1}  \\
OPAL data               & 14             & 13        \\ \hline

cross-section (measured) (pb)& 4.1$\pm$1.6   & 0.9$\pm$0.3   \\ \hline
cross-section (grc4f)    (pb)& 4.21$\pm$0.02 & 0.59$\pm$0.01 \\
cross-section (PYTHIA)   (pb)& 4.04$\pm$0.03 & 0.72$\pm$0.02 \\ \hline
\end{tabular} 
\caption{\it Comparison of the OPAL data with the cross-sections
predicted by the grc4f and the PYTHIA Monte Carlo generators for
$5~\gev < \mqq < 60~\gev$ (``ee$\gamma^*$'' region) and for $\mqq \geq
60~\gev$ (``eeZ'' region). The efficiencies have been obtained with
grc4f.  Only statistical errors are given. The errors on the
feed-through are negligible.}
\label{tab:vglneu} 
\end{center} 
\end{table} 

In Fig.~\ref{fig:plotallcuts} we have plotted after all cuts the same
distributions as in Fig.~\ref{fig:plotpres}. The respective
contributions of the $\gamma^*$ and of the Z to the measured process
are clearly visible in the $\mqq$\ distribution. The $\eelec$\
distribution shows the preference for small electron momenta due to
the low transverse momenta involved in the e$\gamma$ scattering.

\section{Systematic studies}
\label{syserrors}

Several sources of systematic errors have been considered.
Uncertainties in the description of the signal process may cause a
systematic error on the efficiencies. These come mainly from imperfect
modelling of the detector and have been analysed by comparing the
Monte Carlo simulation to real data for the process $\ee\to\ww$\ where
one W decays hadronically and the other one to an electron and a
neutrino. This process has the same observable final state as \Zgee\
and can therefore be used to check for any discrepancies in the
description of the cut variables which would lead to an error in the
selection efficiency. After selecting the $\ee\to\ww\to{\rm q{\bar
q}e^-}{\bar\nu}_{\rm e}$ events using the procedure described in
Ref.~\cite{wwselection}, each cut was applied separately to this
sample. The difference in selection efficiency between data and the
Monte Carlo simulation was taken as the systematic error after
subtracting quadratically the statistical error of the WW sample. In
those cases where the statistical error was larger than the efficiency
differences, we have conservatively taken the statistical error as
systematic uncertainty. The results of the check are shown in
Table~\ref{tab:syseffi}. In our error estimate we have considered the
fact that some distributions of the process $\ee\to\ww\to{\rm q{\bar
q}e^-}{\bar\nu}_{\rm e}$ are significantly different from those of the
signal (e.g. $\qecosqq$\ or $|\costmiss|$). These systematic
uncertainties have been used for both the ``ee$\gamma^*$''-like and
the ``eeZ'''-like hadronic mass region, since the efficiencies of each
selection cut are similar in the two regions. As mentioned above, the
relatively smaller overall efficiency for ``ee$\gamma^*$''-like events
is due to the multiplicity cut in the preselection. To assess the
systematic uncertainty of this cut, we have calculated the change in
efficiency when the simulated $N_{\rm mult}$ distribution is shifted
by $\pm 1$. Half of the resulting change is quoted as uncertainty for
this cut in the first row of Table~\ref{tab:syseffi}.  The total
relative systematic uncertainty on the signal efficiencies due to
event simulation is then found to be 0.086 for high mass and 0.107 for
low mass hadronic systems, respectively.

\begin{table}[htb]
\begin{center} 
\begin{tabular}{|l||c|c|} \hline
&$5~\gev < \mqq < 60~\gev$ & $\mqq \geq 60~\gev$              \\

                        &``ee$\gamma^*$''& ``eeZ''            \\ \hline \hline
$N_{\rm mult}>8$               & 0.063 & 0.000      \\ 
$\qecosqq > 0.75$              & 0.018 & 0.018      \\ 
$\qecose < 0.75$               & 0.037 & 0.037      \\ 
$|\costmiss| > 0.95$           & 0.047 & 0.047      \\ 
$\pmiss > 30$~GeV              & 0.036 & 0.036      \\ 
electron isolation             & 0.046 & 0.046      \\ 
$\Efwd < 35$~GeV               & 0.009 & 0.009      \\ \hline \hline
event simulation               & 0.107 & 0.086      \\ 
generator distributions        & 0.063 & 0.063      \\ 
background                     & 0.070 & 0.018      \\ \hline \hline
Total                          & 0.142 & 0.108      \\ \hline 
\end{tabular} 
\caption{\it 
Relative systematic uncertainties of the cross-section measurements.
The entry ``event simulation'' is the quadratic sum of the signal
efficiency uncertainties for the single cuts listed in the rows above
it.}
\label{tab:syseffi} 
\end{center} 
\end{table} 

A smaller contribution is the uncertainty of the efficiency due to
systematic errors in the Monte Carlo generator distributions. This has
been estimated from the relative difference between the measured
``ee$\gamma^*$'' and ``eeZ'' cross-sections obtained using the
efficiencies from grc4f and those from PYTHIA.  The relative
cross-section differences are very similar in both hadronic mass
regions, and statistically compatible with zero.  Adding the
cross-sections of both regions, a relative difference of
0.033$\pm$0.063 between grc4f and PYTHIA is found. In
Table~\ref{tab:syseffi} we conservatively take the statistical error
on this difference as systematic uncertainty for both the
``ee$\gamma^*$'' and the ``eeZ'' region.

Uncertainties in the description of the background processes cause
systematic errors on the background subtraction. The contributions to
these errors coming from $\eetoqqbar$, two-photon processes and other
four-fermion backgrounds have been tested separately for each
background. By inverting or removing one or two selection cuts while
the others were left unchanged, we produced samples where the analysed
background process is significantly dominant. We obtained the
systematic uncertainty for the various background processes by taking
the relative difference of the total event numbers in data and Monte
Carlo after the statistical error on the data has been subtracted
quadratically. The relative systematic uncertainties we obtain are
0.17 for the $\eetoqqbar$\ background, 0.14 for the four-fermion
backgrounds and 0.24 for the two-photon process $\eetogg$. The numbers
of background events after all cuts are 1.8$\pm$0.5$\pm$0.3,
1.4$\pm$0.1$\pm$0.2 and 3.1$\pm$0.8$\pm$0.7 for $\eetoqqbar$, $\ee\to$
4f and $\eetogg$, respectively. Since other background sources are
much smaller, their systematic uncertainties have been neglected.  The
total error on the background events corresponds to a relative
systematic uncertainty for the obtained cross-sections of 0.070 for
``ee$\gamma^*$'' and 0.018 for ``eeZ'' (see Table~\ref{tab:syseffi}).

\section{Results and conclusion}
\label{results}

Using a data set with an integrated luminosity of
(54.7$\pm$0.1$\pm$0.2)~$\ipb$ collected with the OPAL detector at
$\sqrt{s}=183$~GeV, 27 candidate events for the process
\Zgee\ have been selected, with an expected background of
6.5$\pm$0.9$\pm$0.8 events, where the first errors are statistical and
the second are systematic. The expected number of signal events is
16.7$\pm$0.3$\pm$1.9.

From the observed number of events, the cross-section for this process
has been calculated separately for a high mass, ``eeZ''-like, and a
low mass, ``ee$\gamma^*$''-like region, with a cut at a hadronic mass
of 60~GeV. Within the phase space of our kinematic signal definition
described in Section~\ref{sigdef}, we measure a cross-section of
\csrlo\ for ``ee$\gamma^*$''-like events, and a cross-section of
\csrhi\ for ``eeZ''-like events, including the systematic errors
obtained in Section~\ref{syserrors}.

To investigate in more detail the kinematic properties of the Compton
scattering process, we show in Fig.~\ref{fig:plotstut} the
differential distributions of the Mandelstam variables from
Equation~\ref{eq:compton}, and of the cosine of the scattering angle
$\theta^*$ of the $\Zg$\ in the e$\gamma$ rest system with respect to
the incoming $\gamma$ direction. All quantities have been obtained
from the measured momenta and energies of the electron and the
hadronic system, and are compared with the grc4f signal prediction.
The structure observed in the $\sqrt{\hat{s}}$ distribution is
consistent with expectation for the $\gamma^*$ (low $\sqrt{\hat{s}}$)
and Z (high $\sqrt{\hat{s}}$) contributions. An excited electron,
contributing in the $\hat{s}$-channel of the signal diagram, would
show up as a peak at the e$^*$ mass.  For illustration, we have added
in Fig.~\ref{fig:plotstut}a as a dashed histogram the expected
contribution of events with an excited electron of mass $m_{\rm e}^* =
120~$GeV, normalized to a product of cross-section times branching
ratio of 5~pb. It has been obtained by applying our analysis to fully
simulated events of the process $\ee\to$~ee$^*\to$~eeZ from the Monte
Carlo generator EXOTIC~\cite{PR219}.

In the $\that$\ and $\uhat$\ distributions, peaks near zero are
predicted, with a long tail to large negative values in the case of
$\that$. The measured $\that$\ distribution agrees well with this
prediction and also shows the drop near zero, due to angular
acceptance cuts.  The observed $\uhat$\ distribution seems somewhat
broader than predicted, though it is still consistent within errors.
The entries at positive values are due to detector resolution.  The
distribution of $\cos\theta^*$ peaks at $-1$, corresponding to small
$\hat{u}$ which is typical for Compton processes.

In summary, we have reported the first observation of $\Zg$\
production in Compton scattering of quasi-real photons which is a
subprocess of the reaction \Zgee.  The predictions of the grc4f and
PYTHIA Monte Carlo programs, as listed in Table~\ref{tab:vglneu}, are
both in good agreement with our results.  The data statistics are not
yet sufficient to be sensitive to the difference between the
predictions of the Monte Carlo generators.

\appendix
\section*{Acknowledgements}
We thank T. Sj{\"o}strand for the useful discussions.\\
We particularly wish to thank the SL Division for the efficient operation
of the LEP accelerator at all energies
 and for their continuing close cooperation with
our experimental group.  We thank our colleagues from CEA, DAPNIA/SPP,
CE-Saclay for their efforts over the years on the time-of-flight and trigger
systems which we continue to use.  In addition to the support staff at our own
institutions we are pleased to acknowledge the  \\
Department of Energy, USA, \\
National Science Foundation, USA, \\
Particle Physics and Astronomy Research Council, UK, \\
Natural Sciences and Engineering Research Council, Canada, \\
Israel Science Foundation, administered by the Israel
Academy of Science and Humanities, \\
Minerva Gesellschaft, \\
Benoziyo Center for High Energy Physics,\\
Japanese Ministry of Education, Science and Culture (the
Monbusho) and a grant under the Monbusho International
Science Research Program,\\
German Israeli Bi-national Science Foundation (GIF), \\
Bundesministerium f\"ur Bildung, Wissenschaft,
Forschung und Technologie, Germany, \\
National Research Council of Canada, \\
Research Corporation, USA,\\
Hungarian Foundation for Scientific Research, OTKA T-016660, 
T023793 and OTKA F-023259.\\



\clearpage
\begin{figure}[htb]
\centerline{\epsfig{figure=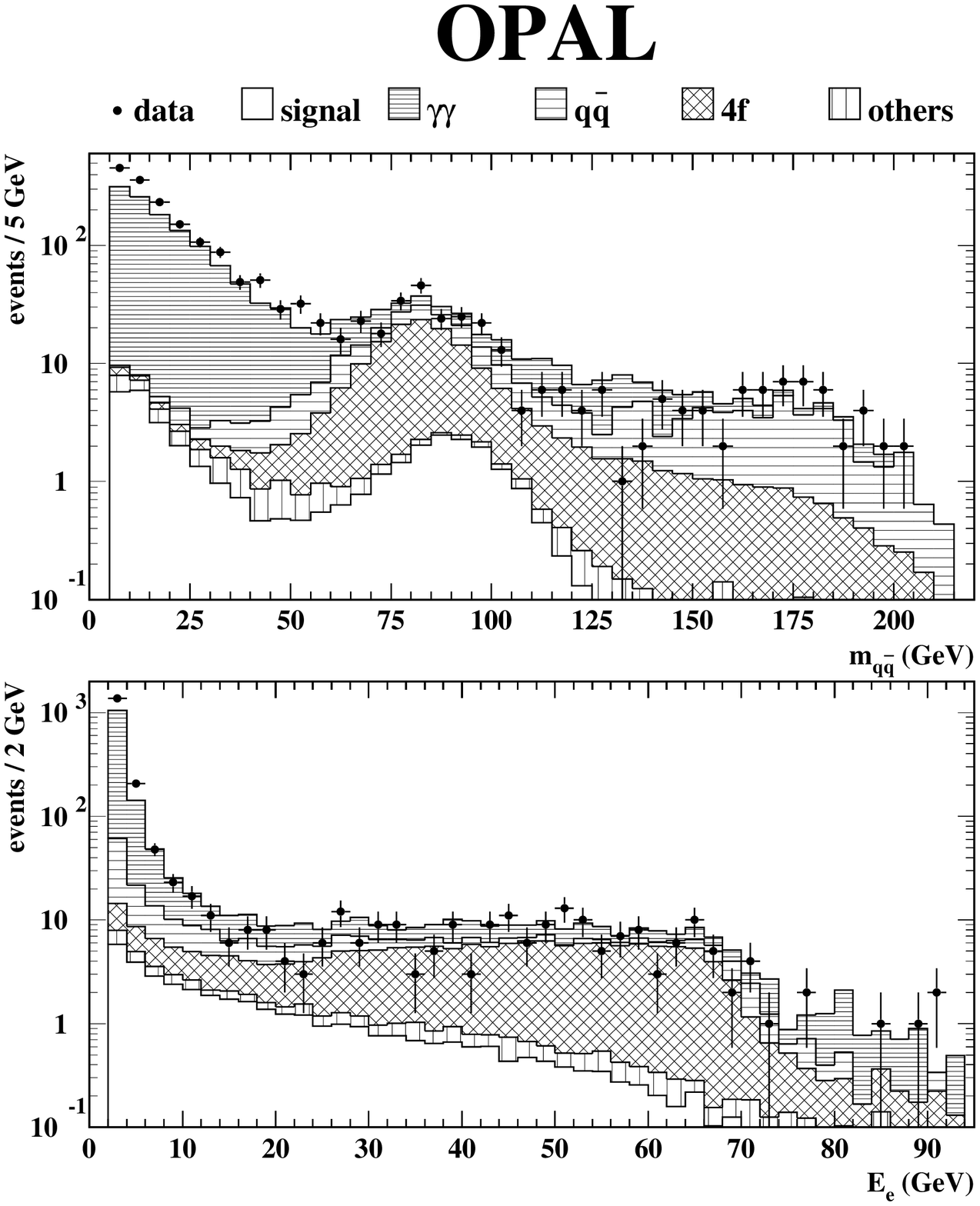,width=1.0\textwidth}}
\caption{\it 
Distributions of $\mqq$\ and $\eelec$\ after the preselection. The
histograms show the contributions from the various Monte Carlo
simulations and the points are the data. The sharp lower edges of the
spectra are due to preselection cuts. Only statistical errors are
shown.}
\label{fig:plotpres} 
\end{figure} 

\clearpage
\begin{figure}[htb]
\centerline{\epsfig{figure=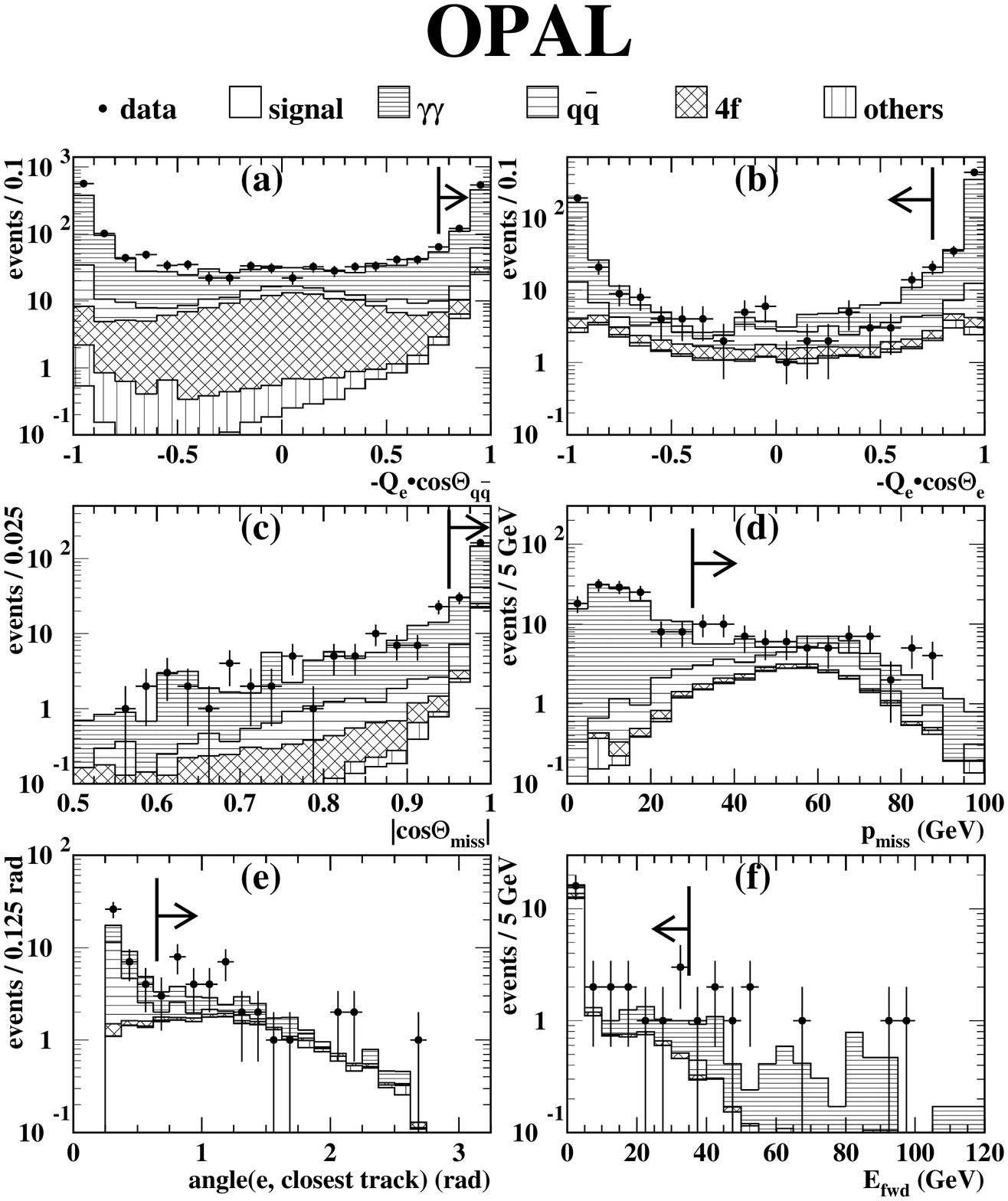,width=1.0\textwidth}}
\caption{\it 
Distributions of the selection variables. The histograms show the
various Monte Carlo simulations and the points are the data. The
arrows point into the selected regions.  The cuts have been applied
successively. Only statistical errors are shown.}
\label{fig:plotcuts} 
\end{figure} 

\clearpage
\begin{figure}[htb]
\centerline{\epsfig{figure=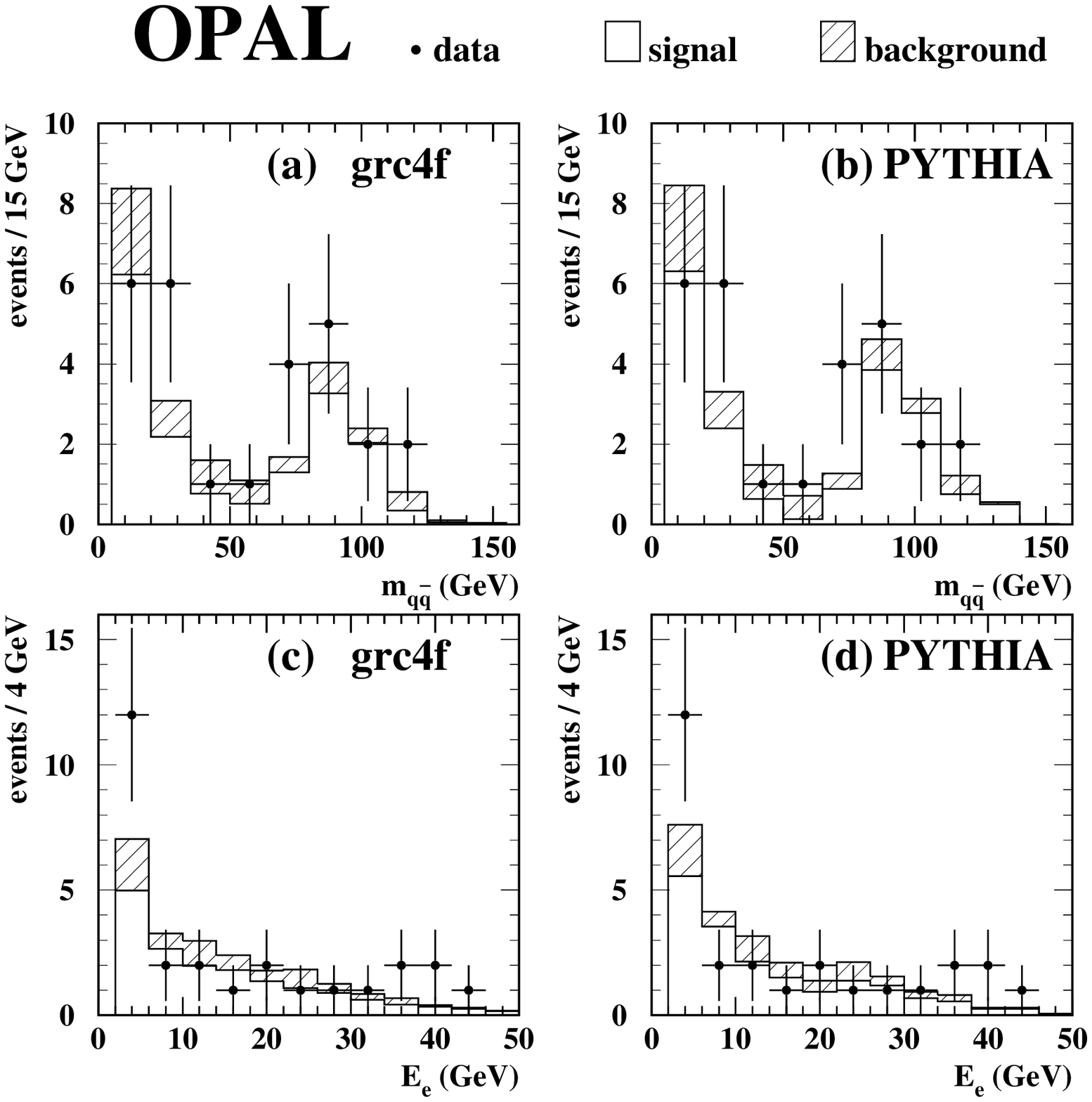,width=1.0\textwidth}}
\caption{\it 
Distributions of $\mqq$\ and $\eelec$\ after all cuts. In Figures~(a)
and (c) the signal is simulated using the grc4f generator. In (b) and
(d) we have used PYTHIA instead of grc4f. The histograms show the
various Monte Carlo simulations and the points are the data. Only
statistical errors are shown.}
\label{fig:plotallcuts} 
\end{figure} 
\clearpage

\begin{figure}[htb]
\centerline{\epsfig{figure=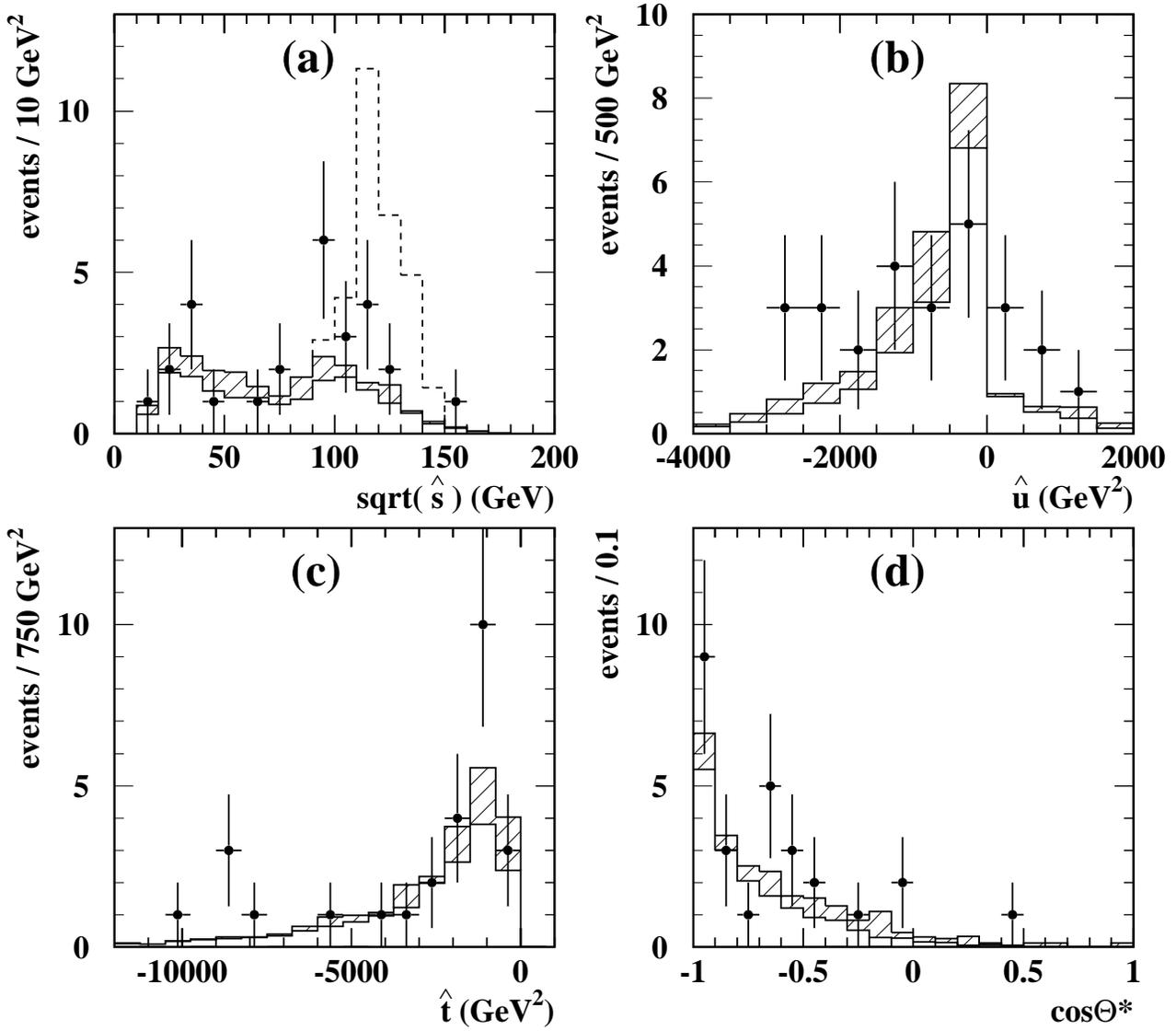,width=1.0\textwidth}}
\caption{\it 
Distributions of $(\shat)^{1/2}$, $\that$, $\uhat$\ and $\cos\theta^*$
after all cuts. The histograms show the various Monte Carlo
simulations and the points are the data. The dashed histogram in (a)
is a hypothetical e$^*$ signal, as described in the text. Only
statistical errors are shown.}
\label{fig:plotstut} 
\end{figure} 

\end{document}